\documentclass{cs20proc}

\editors{S. J. Wolk}
\publisher{Zenodo}
\conference{The 20th Cambridge Workshop on Cool Stars, Stellar Systems, and the Sun}
\conferencedate{2018}

\title{Chemical connections between low-mass stars and planet building blocks investigated by stellar population synthesis}
\author{Cabral Nahuel,$^{1}$ 
        Nad\`ege Lagarde,$^{1}$
        C\'eline Reyl\'e,$^{1}$
        Aur\'elie Guilbert-Lepoutre$^{1}$        
        and Annie C. Robin$^{1}$        
        }

\affiliation{$^{1}$ Institut UTINAM, CNRS UMR6213, Univ. Bourgogne Franche-Comt\'e, OSU THETA Franche-Comt\'e-Bourgogne, Observatoire de Besan\c con, France}

\shorttitle{Ramblings of Kant}
\shortauthors{Nahuel Cabral}

\abs{Connecting star and planet properties in a single model is not straightforward. Stellar population synthesis models are key to explore combined statistical constraints from star and planet observations. The Besan\c{c}on stellar population synthesis model \citep[e.g.][]{Robin2003, Lagarde2017} includes now the stellar evolutionary tracks computed with the stellar evolution code STAREVOL \citep[e.g.][]{Lagarde2012, Amard2016}. It provides the global (M, R, Teff, etc) and chemical properties of stars for 54 chemical species. It enables to study the different galactic populations of the Milky Way (the halo, the bulge, the thin and thick disc) and a specific observational survey. Here, we couple the Besan\c{c}on model with a simple stoichiometric model \citep[][]{Santos2017} in order to determine the expected composition of the planet building blocks (PBB). We investigate the trends and correlations of the expected chemical abundances of PBB in the different stellar populations of the Milky Way \citep{Cabral2018}.}

\begin{document}

\maketitle

\section{Introduction}
%\kant Now, for something really huge. I mean, really huge. Like, huger than huge can be. We're going to end a sentence with a footnote.

Future spatial missions (TESS, CHEOPS, PLATO and JWST) will improve considerably our understanding of the formation of planetary systems. Exploring connections between star and planet properties provide key constrains on planet formation models. Currently, space-based (Kepler, CoRoT) and ground-based exoplanet surveys (HARPS) have shown that the presence of planetary companions is closely linked to the metallicity and the chemical abundances of the host stars \citep[e.g.][]{Adibekyan2015}.  Moreover, TESS, CHEOPS, and JWST are expected to bring huge improvements in our characterization of planets. The asteroseismic survey PLATO will provide stellar properties with high accuracy \citep[][]{Rauer2014,Miglio2017}.

Planet properties are observed to correlate with metallicity and specific elemental ratios of the host stars. Moreover, the different stellar populations in our Galaxy are characterized by different metallicities and alpha abundances due to their different way and epoch of formation and chemical evolution \citep[][]{Haywood2013}. Thus, the different stellar galactic populations could produce planets with very different properties.
%he halo contains the more metal poor stars. The disk exhibits two sequences (thick and thin discs) where the thick disk stars are in general more metal-poor and alpha-enriched compared to thin disk stars . The bulge presents a range in metallicities similar to the thin disk but a much larger spread in alpha abundances

%Santos et al. (2017, hereafter S17) determined the expected chemical composition of planet building blocks (PBB) in the different galactic populations, using the elemental abundances determined with the HARPS survey (Adibekyan et al. 2012b). 

Stellar synthetic models are particularly useful to analyse star-planet correlations in the Milky Way. In line with the previous work of \citet[][]{Santos2017}, we explore statistical trends of the PBB composition in the different galactic populations. We use the Besan\c{c}on Galaxy Model (BGM) to simulate the global and chemical properties of stars in different populations \citep[][]{Lagarde2017}. We then apply the stoichiometric model from Santos et al. 2017 to these synthetic populations. This allows us to give robust predictions of the PBB for the whole Milky Way. The development of such a coherent and integrated model is important to prepare the interpretation of future large scale surveys of exoplanets search.

\section{Numerical model}

\subsection{The Besan\c{c}on Galaxy Model}

The Besan\c{c}on stellar populations synthesis model provides the global (e.g., M, R, Teff) and chemical properties of stars for 54 chemical species. Four populations are considered: the halo, the bulge, the thin and thick disks. For each one we assume different initial mass functions (IMF) and different formation and evolution histories. In each case, the IMF is a three-slopes power law. The star formation history (SFH) and the IMF of the thick disk and halo have been set from comparisons to photometric data from 2MASS and SDSS surveys \citep[][]{Robin2014}. The star formation rate, in the thin disk population, is assumed decreasing exponentially with time \citep[][]{AumerBinney2009}. The parameters of the SFH and the IMF of the thin disk have been fit to the Tycho-2 catalog (Czekaj et al. 2014).

The new version of the BGM includes a new grid of stellar evolution models computed with the stellar evolution code STAREVOL \citep[e.g.][]{Lagarde2012, Amard2016} for stars with $M \ge$ 0.7 M$_\odot$ \citet[][]{Lagarde2017} and \citet[][]{Lagarde2018}. These stellar evolution tracks have been computed from the pre-main sequence to the early asymptotic giant branch at six metallicities ([Fe/H]= 0.51, 0, -0.23, -0.54, -1.2, -2.14), and at different  $\alpha$-enhancements ([$\alpha$/Fe]=0.0, 0.15 and 0.30) to simulate all populations.

To determine the [$\alpha$/Fe]-[Fe/H] trend, for the four galactic populations, the data release 12 of the APOGEE spectroscopic survey \citep{Majewski2017} has been used : 

\begin{equation}
      \mbox{[$\alpha$/Fe]}=
     \left\{
     \begin{array}{lll}
     0.014+0.01406\times\mbox{[Fe/H]} \\
\hspace{0.9cm}  +0.1013\times\mbox{[Fe/H]}^{2} & \mbox{Thin disk, [Fe/H]<0.1} \\
    0 & \mbox{Thin disk, [Fe/H]>0.1}\\
   %&   \textbf{for thin disk stars, }\\
     0.320-\mbox{e}^{(1.19375\times\rm{[Fe/H]}-1.6)} & \mbox{Thick disk and bulge} \\
   0.3 & \mbox{Halo} \\
     \end{array}
     \right.
\end{equation}

\begin{figure}[h]
  \centering
    \includegraphics[width=\hsize,clip=true,trim= 0.1cm 0.1cm 0.1cm 0.1cm]{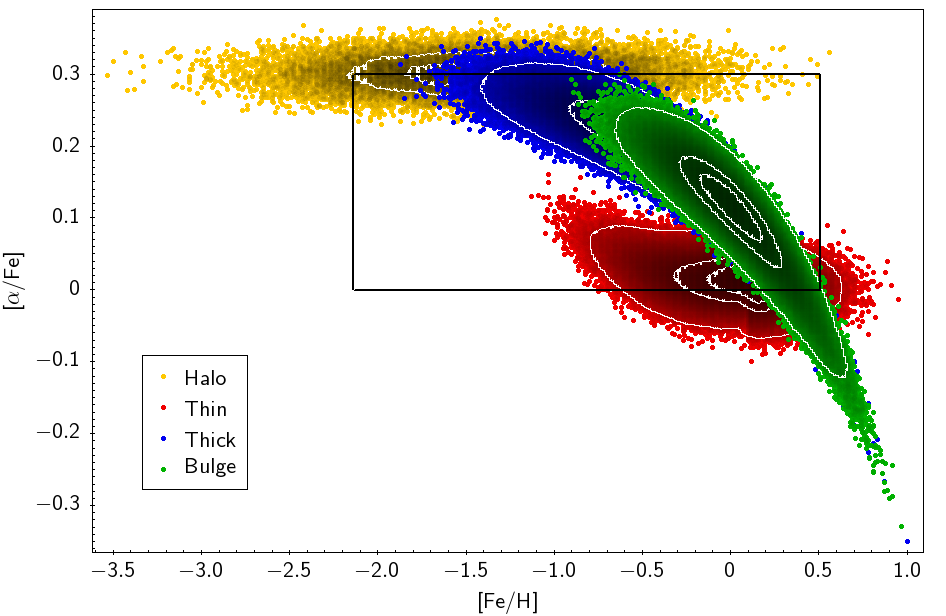}
\caption{The [$\alpha$/Fe] abundance as a function of [Fe/H] for stars with d<50 kpc simulated with the BGM. Thin and thick disks as well as bulge and halo are represented by red, blue, green and yellow dots respectively. Black lines indicate the selected population used in this study: -2.14<[Fe/H]<0.51 and 0<[$\alpha$/Fe]<0.3.}
    \label{FigAlpha}
\end{figure}

An intrinsic Gaussian dispersion of 0.02 dex is added to these relations. Moreover, we necessarily limit our calculations to the border of the stellar grid in metallicity and alpha content -2.14<[Fe/H]<0.51 and 0<[$\alpha$/Fe]<0.3 (see Fig. \ref{FigAlpha}).

\subsection{Stoichiometric model}

We use the stoichiometric model published in \citet{Santos2017}. In this simple model, the molecular abundances in the protoplanetary disk, and their mass fraction, can be computed from the stellar abundances of a handful of elements. Fe, Si, Mg, O and C together with H and He control the species expected from the equilibrium condensation of H$_{2}$, He, CH$_{4}$, H$_{2}$O, Fe, MgSiO$_{3}$, Mg$_{2}$SiO$_{4}$ and SiO$_{2}$ \citep{Lodders2003, Bond2010}. These compounds dominate the rocky interior of Earth-like planet \citep[see e.g.][]{Sotin2007}.

Since we limited our synthetic population to the borders of the stellar grids (-2.14<[Fe/H]<0.51, 0<[$\alpha$/Fe]<0.3) all simulated stars have 1<Mg/Si<2. We write here the inverted stoichiometric relations corresponding to the case 1<Mg/Si<2 (assuming the equations on Appendix B of S17):

\begin{equation}
     \begin{array}{lll}
     N_{MgSiO_{3}} = 2 N_{Si} - N_{Mg} \\
     N_{Mg_{2}SiO_{4}} = N_{Mg} - N_{Si} \\
     N_{H_{2}O} = N_{O} - 2 N_{Si} - N_{Mg} \\
     N_{CH_{4}} = N_C 
     \end{array}
\end{equation}

with $N_X$ the number of atoms of each specie $X$. 

These relations enable the computation of the expected mass fractions of PBB, the iron-to-silicate mass fraction ($f_{iron}$) and the water mass fraction ($f_{w}$) :

 \begin{equation}
     \begin{array}{lllc}
  f_{iron} = m_{Fe}/(m_{Fe}  + m_{MgSiO_3} + m_{Mg_2SiO_4} + m_{SiO_2}) \\
  f_{w} = m_{H_2O} /(m_{H_2O}+ m_{Fe}      + m_{MgSiO_3}   + m_{Mg_2SiO_4} + m_{SiO_2})\\
     \end{array}
 \label{EqFractMass}
 \end{equation}

where $m_{X}=N_{X} . \mu_{X}$ with $N_{X}$ the number of atoms of each species $X$ and $\mu_{X}$ their mean molecular weights. The total mass is given by $M_{tot}=N_{H} . \mu_{H} + N_{He} . \mu_{He} +N_{C} . \mu_{C} + N_{O} . \mu_{O} + N_{Mg} . \mu_{Mg} + N_{Si} . \mu_{Si} + N_{Fe} . \mu_{Fe}$.

\section{Chemical trends of PBB in the Milky Way}

We simulate the FGK sample to distances up to 50 kpc from the Sun. This large volume covers the Milky Way up to the external parts. Our simulation has a total of 4 850 600 000 stars: 42\% from the thin disk, 27\% from the thick disk, 30\% from the bulge, and 1\% from the halo.

\subsection{The water/iron valley}

\begin{figure}
  \centering
    \includegraphics[width=\hsize,clip=true,trim= 0cm 0cm 0cm 0cm]{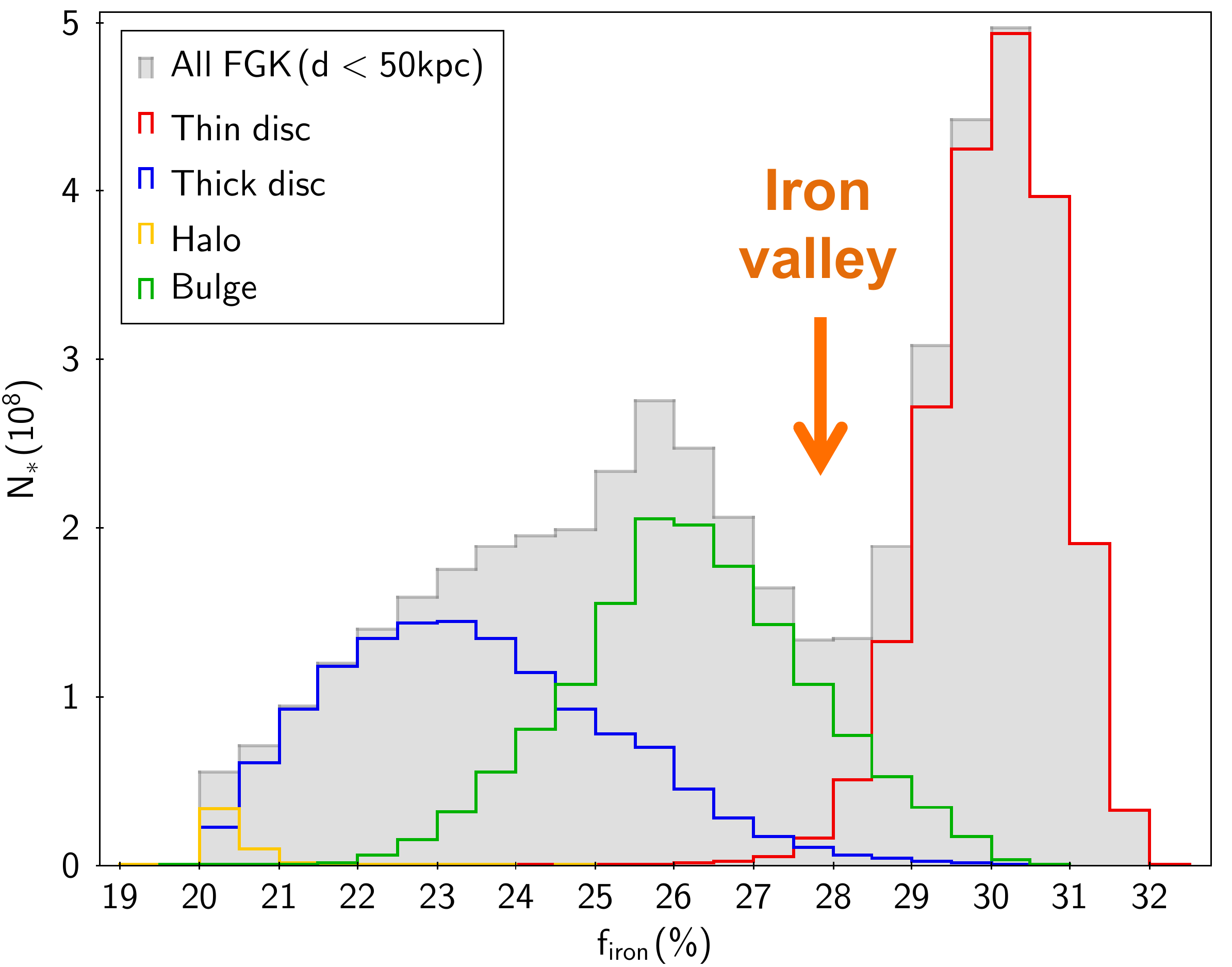}
    \vskip 1cm
    \includegraphics[width=\hsize,clip=true,trim= 0cm 0cm 0cm 0cm]{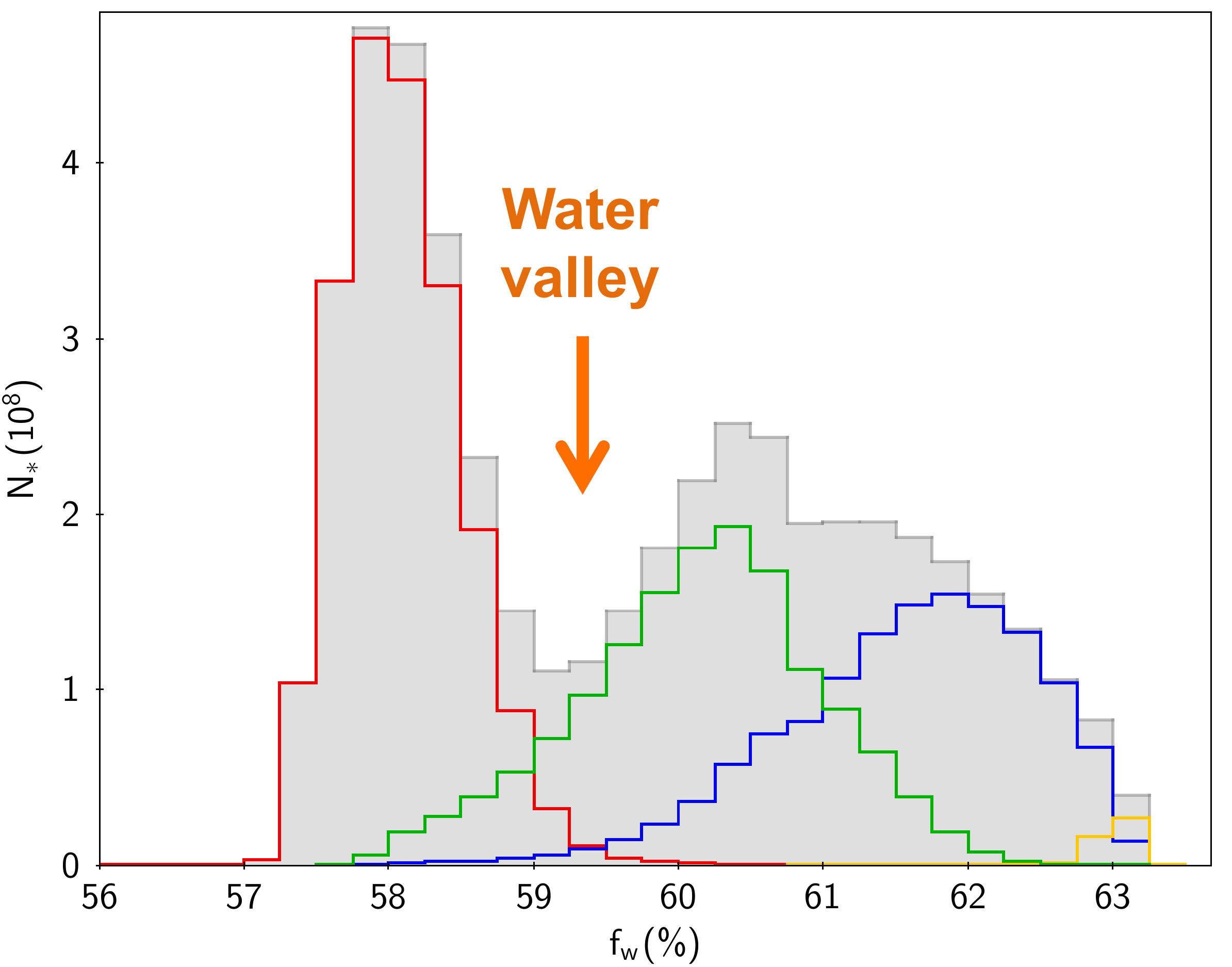}
%    \vskip 1cm
%	\includegraphics[width=\hsize,clip=true,trim= 0cm 0cm 0cm 0cm]{example/MW_fz.pdf}
     \caption{Mass fraction distributions, $f_{iron}$ (upper panel) and $f_{w}$ (middle panel) for the four stellar populations of the Milky Way: thin disk, thick disk, halo and bulge. We ran the model up to distances of 50 kpc.}
    \label{FigMW50Histo}
\end{figure}

\begin{figure}
  \centering
    \includegraphics[width=\hsize,clip=true,trim= 0cm 2cm 0cm 0cm]{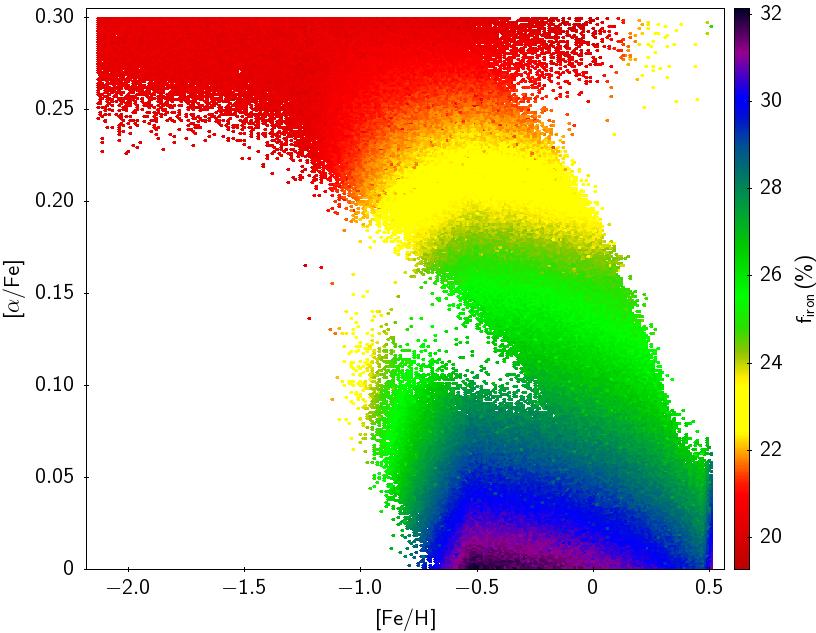}
    \includegraphics[width=\hsize,clip=true,trim= 0cm 0cm 0cm 0cm]{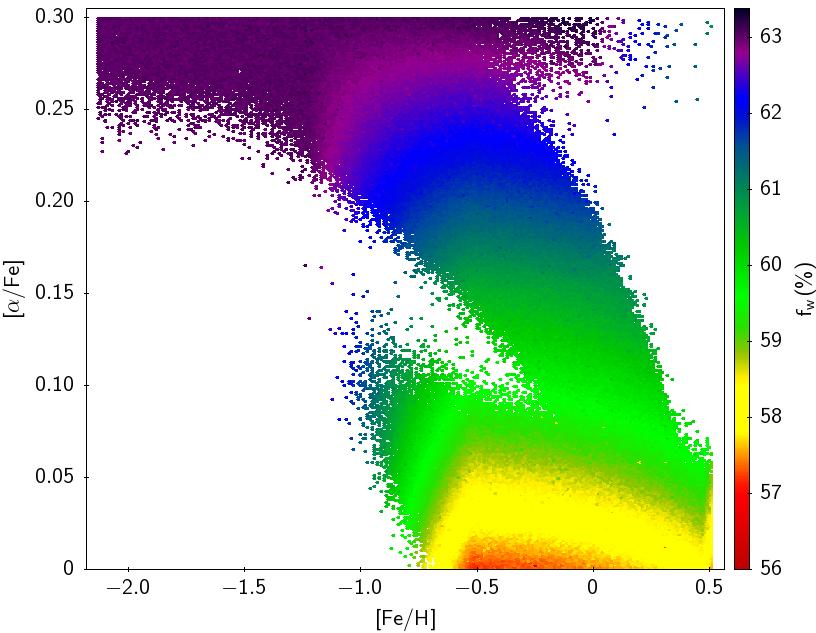}
\caption{The iron-to-silicate mass fraction, $f_{iron}$ (upper panel), and the water mass fraction, $f_{w}$ (bottom panel) for the four stellar populations of the Milky Way: thin disk, thick disk, halo and bulge. We ran the model up to distances of 50 kpc.}
    \label{Fig_Alpha_firon}
\end{figure}

The synthetic population computed with the BGM predicts a distinct distribution between the thin disk stars with iron-rich PBB, and other stellar populations with water-rich PBB, implying a significant dip in the number of stars around $f_{iron}$$\sim$28\% and $f_{w}$$\sim$59\% (Fig. \ref{FigMW50Histo}). This "iron/water valley'' results from the stellar alpha content distributions in the synthetic stellar populations. Indeed, in Fig. \ref{Fig_Alpha_firon} we show that $f_{iron}$ and $f_{w}$ are mainly function of [$\alpha$/Fe]. Moreover, the density of stars around solar metallicity and [$\alpha$/Fe]$\sim$0.1 is smaller due to the known gap between the thin and the thick disk. This gap translates into a bimodal distribution in the $f_{iron}$ and $f_{w}$ histograms. The synthetic population computed with the BGM shows the presence of the "iron/water valley'' because of the clear dependence of $f_{iron}$ and $f_{w}$ on [$\alpha$/Fe] (Fig. \ref{Fig_Alpha_firon}), and because of the known gap between the thin and the thick disk on [$\alpha$/Fe].

% 
% 
% \begin{table}[t]
% 	\centering
% 	\caption{Simple table for testing tabref.}
% 	\label{tab:table_narrow}
% 	\begin{tabular*}{0.85\linewidth}{l @{\extracolsep{\fill}} c l}
% 	\noalign{\smallskip}\hline\hline\noalign{\smallskip}
% 	Star & Data & Source \\
% 	\noalign{\smallskip}\hline\noalign{\smallskip}
% 	Star 1 &  0.5 & \citet{Salpeter1955}  \\
% 	Star 2 &  0.4 & \citet{Allard2011}  \\
% 	Star 3 &  0.3 & \citet{Press1974}  \\
% 	Star 4 &  0.2 & \citet{Riess1998}  \\
% 	Star 5 &  0.1 & \citet{Henyey1959} \\
% 	\noalign{\smallskip}\hline
% 	\end{tabular*}
% \end{table}
% 
% \begin{table*}[t]
% 	\centering
% 	\caption{Simple table for testing tabref.}
% 	\label{tab:table_wide}
% 	\begin{tabular*}{\linewidth}{l @{\extracolsep{\fill}} c l c l}
% 	\noalign{\smallskip}\hline\hline\noalign{\smallskip}
% 	Star & Data & Source & Extra Data & Source \\
% 	\noalign{\smallskip}\hline\noalign{\smallskip}
% 	Star 1 &  0.5 & \citet{Salpeter1955} &  -1.0  & \citet{Salpeter1955, Henyey1959, Press1974} \\
% 	Star 2 &  0.4 & \citet{Allard2011} &  -2.0  & \citet{Salpeter1955, Press1974}  \\
% 	Star 3 &  0.3 & \citet{Press1974} &  -3.0  & \citet{Allard2011, Riess1998}  \\
% 	Star 4 &  0.2 & \citet{Riess1998} &  -4.0  & \citet{Riess1998, Henyey1959}  \\
% 	Star 5 &  0.1 & \citet{Henyey1959} &  -5.0  & \citet{Press1974} \\
% 	\noalign{\smallskip}\hline
% 	\end{tabular*}
% \end{table*}

\subsection{Expected exoplanets composition}

From the expected chemical composition of planet building blocks we can discuss the expected exoplanets composition. The stoichiometric model \citep{Santos2017} predicts for the Sun $f_{iron}$$\sim $$33\%$, $f_{w}$$\sim$$60\%$ and $f_{Z}$$\sim$$1.3\%$. These values of $f_{w}$ and $f_{Z}$ are in line with the values of $f_{w}$=67.11\% and $f_{Z}$=1.31\% found by \citet{Lodders2003}. Moreover, $f_{iron}$$\sim $$3\%$ is consistent with the values observed in the meteorites and the rocky planets of the solar system, Earth, Venus and Mars \citep[see e.g.][]{DrakeRighter2002}.
%However, terrestrial planets of the solar system present a large spread in the water content. As discussed in S17, the value of water mass fraction is expected to present strong scatter compared to the values determined with the simple stoichiometric model used in this study. Indeed, several physical processes can modify the water content of the rocky planets. For instance, the origin of water on Earth could be due to comets and asteroids infall. On the other hand, the migratory path and the initial location of the embryos could be crucial in the final water content, depending on where the iceline is located. The simple model we use here neglects this kind of mechanisms. The values of $f_\text{w}$ should be analysed with caution. They should be only indicative of the water content in the protoplanetary disc. In spite of these considerations, the water content of the protoplanetary disc should probably correlate with the final water content of planetesimals.

Figure \ref{FigMW50Histo} shows that the synthetic thin disk stars present PBB chemical composition compatible with the values of the solar system. Thus, the thin disk stars could produce planets with similar composition to the rocky planets of the solar system. Metal-poor stars of the thick disk might produce planets with lower iron mass. Similar results are obtained by \cite{Santos2017} with HARPS observations. This should have an impact on the radius of rocky planets. Indeed, a low iron mass fraction may produce a smaller core \citep{Valencia2007}. However, statistically the thick disc stars should present much higher water mass fractions. It is worth noting that, the bulge presents PBB compositions similar to the solar ones, and should also be able to host rocky planets with solar-system-like composition. Instead, halo stars should produce water rich planets.

%When we compare the solar values with the values obtained by the MW simulation, we observe that the synthetic thin disc stars present PBB chemical composition compatible with the values of the solar system. As expected, stars of the thin disc could produce planets with similar composition to the rocky planets of the solar system. Metal-poor stars of the thick disc might produce planets with lower iron mass.

% Currently there is only one planetary mass object detected in the bulge, the brown dwarf OGLE-2017-BLG-1522Lb \citep{Jung2018} and there is no planet detected orbiting halo stars. 

\section{Conclusion}

The different stellar populations we observe in our Galaxy are characterized by different metallicities and alpha-element abundances. We aim to build an integrated tool to predict the planet building blocks composition as a function of the stellar populations. We investigate the trends of the expected PBB composition with the chemical abundance of the host star, in different parts of the Galaxy, predicted by stellar population synthesis. The new version of the BGM, including the simple stoichiometric model derived by S17, appears as a powerful tool to predict chemical composition of PBB. This may be crucial to prepare interpretation of on-going and future large scale surveys of exoplanet search.

We run the BGM to generate a synthetic sample of FGK stars up to 50 kpc. This enables to establish general trends for our Galaxy. The main results obtained in this work:

\begin{itemize}  

\item \textsf{Alpha content dependence:} Iron and water mass fractions, $f_{iron}$ and $f_{w}$, are mainly function upon the initial alpha content [$\alpha$/Fe] of the host stars. We have shown that this dependence explains well the iron/water valley.\\

\item \textsf{Iron/water valley:} Since it exists a gap on [$\alpha$/Fe] between the thin disk and the other galactic populations (thick disk, bulge and halo), and because $f_{iron}$ and $f_{w}$ have a clear dependance on [$\alpha$/Fe], our simulations show an iron/water valley. The valley is predicted by our simulations to be at $f_{iron}$$\sim$$28\%$ and $f_{w}$$\sim$$59\%$. \\

%The different galactic populations (thin and thick disc, halo and bulge) present different metallicities and alpha content [$\alpha$/Fe] distribution, and because the clear dependence on [$\alpha$/Fe] leads to a bimodal distribution of $f_{iron}$ and $f_{w}$ histograms between the thin disk stars and the other galactic populations: the iron/water valley. The valley is predicted by our simulations to be at $f_{iron}$$\sim$$28\%$ and $f_{w}$$\sim$$59\%$. \\

\item \textsf{PBB trends:} Our simulations show that the thin disk is expected to present iron rich and water poor PBB, while the thick disk should have iron poor but water rich PBB. Santos et al. 2017 found similar results with the HARPS observations. \cite{Cabral2018}, with simulations of the solar neighborhood (d<100pc and d<1kpc) confirm these general trends. The bulge presents intermediate values of $f_{iron}$ and $f_{w}$, between the ones of the thin and thick discs. Since its mass fraction values overlap with the ones of the thin disc, it could produce rocky planets with solar-system-like composition. \\ %Instead, the halo stars should produce water-rich planets with low values of iron mass fraction. \\

%\item \textsf{Expected rocky planet trends:} Planet internal structure models predict smaller radii when the iron mass fraction increases and higher radii when water mass fraction increases \citep{Valencia2007}. However, the impact of water mass fraction is clearly dominant when determining the planetary radius. Thus, our trends PBB composition may suggest larger planet radii for the thick disc and the halo stars than for the bulge and the thin disc stars. On the other hand, local processes (we do not take into account in our simple model) could sensibly impact the water mass fraction values. However, albeit complex processes (as comet infall and body migration) are not taken into account here, the final mass fraction values of planets and PBB should correlate with values determined here. In other words, the values of mass fractions predicted here could differ from the real ones, but trends derived by the model should be realistic.\\

\end{itemize}

We linked host star abundances and expected PBB composition in an integrated model of the Galaxy \citep{Cabral2018}. Derived trends are an important step for statistical analyses of expected planet properties. In particular, internal structure models may use these results to derive statistical trends of rocky planets properties, constrain habitability and prepare interpretation of on-going and future large scale surveys of exoplanet search.

%\section*{Acknowledgments}
%{Thank you to the world and myself. Mostly myself. Without me, I would have no results and no paper.}

\bibliographystyle{cs20proc}
\bibliography{example.bib}

\end{document}